\documentstyle[12pt]{article}

\begin{document}
\topmargin 0pt
\oddsidemargin 7mm
\headheight 0pt
\topskip 0mm

\addtolength{\baselineskip}{0.40\baselineskip}

\hfill KAIST-CHEP-96/03

\hfill February 1996

\begin{center}

\vspace{36pt}
{\large \bf $R_b$-$R_c$ Crisis and the Higgs Boson Mass from LEP Precision Data}

\end{center}

\vspace{36pt}

\begin{center}

Jae Sik Lee
\footnote{e-mail address: jslee@chep6.kaist.ac.kr} and Jae Kwan Kim

\vspace{20pt}

{\it Department of Physics,  \\
Korea Advanced Institute of Science and Technology, \\
Taejon 305-701, Korea } \\

\end{center}

\vspace{10pt}

\vfill

\begin{center}
{\bf ABSTRACT}
\end{center}

We study the effects on the Higgs boson mass from LEP precision data of the
new physics explaining $R_b$-$R_c$ crisis.
We implement a fit to LEP observables with the new physics.
We obtain $M_H^{\rm New Physics} = 85 ^{+ 467}_{-56} ~~{\rm GeV}$. 
Comparing with the value of the SM fit,
$M_H^{\rm SM} = 38^{+ 96}_{-21} ~~{\rm GeV}$, the errors are larger and
the central value is higher. The new physics may allow $M_H$ to have a
value out of the range of ${\cal O}(M_Z)$.

\vspace{12pt}

\begin{flushleft}
\end{flushleft}

\noindent

\vspace{24pt}

\vfill

\vspace{15pt}

\newpage

\newpage

It is remarkable that the top-quark mass $M_t$ measured at CDF and D0 agrees well
with the value predicted by the LEP precision data [1]. The success of the the
$M_t$ prediction shifts the focus of interest to the prediction of
the Higgs boson mass $M_H$ [2]. It is shown that there is a weak preference 
for a light Higgs boson mass $M_H < 300$ GeV. But it is not trivial that the
electroweak data have consistently favored a Higgs mass in a range of 
${\cal O}(M_Z)$ [3].

Recently it was reported by LEP collaborations that the measured ratios of 
$R_b \equiv \Gamma(Z\rightarrow b\bar{b})/\Gamma(Z \rightarrow {\rm hadrons})$ 
and $R_c \equiv \Gamma(Z\rightarrow c\bar{c})/\Gamma(Z \rightarrow {\rm hadrons})$ 
are different from those predicted by the standard model (SM). $R_b$ is higher than
the SM prediction at 3.7 $\sigma$ level and $R_c$ is smaller than that at 2.7 
$\sigma$ level [1].
These discrepancies may be the first signals for new physics beyond the SM if these
are confirmed by future measurements.

A number of possible scenarios of new physics are being suggested to explain
these $R_b$ and $R_c$ discrepancies simultaneously [4,5].
The nonuniversal interactions acting on only the b-quark and c-quark are attractive
candidates for new physics explaining these discrepancies since the SM predictions
for other flavors should not be disrupted by the new physics [6]. But it is not possible
to explain $R_b$, $R_c$ with consistent $\alpha_s$ from low energy determinations
invoking only non-standard $Zbb-$ and $Zcc-$couplings. With only nonuniversal interactions
acting on the b-quark and c-quark results in $\alpha_s = 0.18$ [5]. This value is significantly
conflict with the low energy determination $\alpha_s = 0.112 \pm 0.005$ [7]. 
If we don't discount the measured value of $R_c$, therefore,
new physics corrections to the $Zss-$couplings are also needed.

In this paper, we study the effects of the new physics which are introduced to explain
$R_b$ and $R_c$ discrepancies on the Higgs boson mass prediction from the
LEP precision data. By $\chi^2$ fitting to the LEP observables 
we calculate the new physics scale of the nonuniversal interactions and 
obtain $M_H$. There are theoretical bounds on the SM Higgs boson mass which are
obtained from the stability of the electroweak vacuum [8] and 
by requiring the SM couplings to remain perturbative up to some scale [9].
We briefly comment whether our results of fitting are compatible with those from
the vacuum stability and perturbativity.

In this paper we do not construct a specific model but use the
effective Lagrangian technique. We take the $Z \rightarrow f\bar{f}$ vertex to
be given phenomenologically by the expression
\begin{equation}
{\cal L} \sim Z^{\mu} \left[ \bar{f}\gamma_{\mu}(g_V^{{\rm eff},f}+ \right.
\left. g_A^{{\rm eff},f}\gamma_5)f\right],
\end{equation}
where $g_V^{{\rm eff},f}$ and $g_A^{{\rm eff},f}$ are the effective vector
and axial vector coupling constants given by
\begin{eqnarray}
g_V^{{\rm eff},f}&=&2(g_L^{{\rm eff},f}+g_R^{{\rm eff},f}), \nonumber \\
g_A^{{\rm eff},f}&=&2(g_L^{{\rm eff},f}-g_R^{{\rm eff},f}). 
\end{eqnarray} 
We introduce the nonuniversal interactions for $f=s,c,b$.
For $f=c,b$, we parametrize the nonuniversal interaction effects in the
$Z \rightarrow f\bar{f}$ vertex by introducing the parameters
$\kappa_{L,R}^f$. These parameters shift the SM tree level 
couplings of the neutral currents $g_{L,R}^f$ to the effective couplings 
$g_{L,R}^{{\rm eff},f}$:
\begin{equation}
g_{L,R}^{{\rm eff},f} = g_{L,R}^f(1+\kappa_{L,R}^f),
\end{equation}
where
\begin{equation}
g_{L}^f=I_3^f-Q^f\sin^2\theta_W,~~
g_{R}^f=-Q^f\sin^2\theta_W.
\end{equation}
$I_3^f$ and $Q^f$ are the weak isospin and electric
charge respectively. Since non-standard couplings to the strange quark
enter the neutral current observables only via their contributions to the
total hadronic width of the $Z^0$ boson, $\Gamma_{\rm had}$, we parametrize
the effects by introducing the parameter $\delta \Gamma_s$. It is expected that
the $\delta \Gamma_s$ is positive and has the value which nearly cancels the 
deficit of $\Gamma_c$ [7].
Since $(g_L^f)^ 2 \gg (g_R^f)^2$, we fix $\kappa_R^{c,b}=0$ in our
analysis. So we introduced three parameters of new physics : $\kappa_L^b $,
$\kappa_L^c $ and  $\delta \Gamma_s $.

We use the following set of 15 variables in our fitting procedure (see Table 1) :
$\Gamma_Z$, $\sigma_{\rm tot}$, $R_e \equiv \Gamma_{\rm had}/\Gamma_e$,
$R_{\mu} \equiv \Gamma_{\rm had}/\Gamma_{\mu}$,
$R_{\tau} \equiv \Gamma_{\rm had}/\Gamma_{\tau}$, $A_{\rm FB}^0(e)$,
$A_{\rm FB}^0(\mu)$, $A_{\rm FB}^0(\tau)$, $A_{\tau}$, $A_e$, $R_b$,
$R_c$, $A_{\rm FB}^0(b)$, $A_{\rm FB}^0(c)$ and $\sin^2\theta_W^{\rm lep}$.
From Table 1, we can see there are three observables which show deviations
from the predictions of the SM : $A_{\rm FB}^0(\tau)$, $R_b$ and $R_c$.
The inability of the effective Lagrangian approach to fully explain the 
deviations in the asymmetry observables is discussed in Ref. [5]. And
it is out of the range of this paper to consider $A_{\rm FB}^0(\tau)$ deviation
as the effect of the new physics. So we regard that this deviation results from
not well understood systematic effects in the experiments.

We fix $\alpha_s=0.123$ since the strong coupling constant is no longer strongly
constrained by fits with the new physics [5]. We take another 
value of $\alpha_s=0.112$ from low energy
determinations to investigate the effects of the procedure of fixing $\alpha_s$
in our fit. We observe that the effects of varying $\alpha_s$ are negligible.

We used ZFITTER [10] with the function minimizing program MINUIT [11] to perform
the $\chi^2$ fit for the LEP observables. Firstly, we implement the SM fit where
no new physics parameters are added. And we fix $M_H=300$ GeV to see the reliability
for subsequent fits. In this case, the
fitting parameters are $M_t$ and $\alpha_s$. We obtain
\begin{eqnarray}
M_t &=& 171.5 \pm 8.4 ~~{\rm GeV}, \nonumber \\
\alpha_s &=& 0.123 \pm 0.004. \nonumber 
\end{eqnarray}
These values are well agree with those reported by the LEP electroweak 
working group [1].
Note the agreement of the fitted value of $M_t$ with the value measured 
at CDF and D0 : 180 $\pm$ 12 GeV (CDF + D0) [12].

Next, we implement the SM fit where no new physics parameters are added.
In this case we fix $\alpha_s=0.123$. Fixing $\alpha_s$ is for comparisons 
with the results from subsequent fits including
new physics parameters. 
In this case, the fitting parameters are $M_t$ and
$M_H$. We obtain
\begin{eqnarray}
M_H &=& 38^{+ 96}_{-21} ~~{\rm GeV} 
\left[ \log_{10}(M_H)=1.53^{+0.60}_{-0.30} \right], \nonumber \\
M_t &=& 145.3^{+ 16.7}_{- 11.4}. \nonumber 
\end{eqnarray}
The lower and upper errors are obtained by projecting the $\Delta \chi^2 =1$ 
ellipse in $(M_t,\log_{10}(M_H))$ plane on the vertical and horizontal axes.
$M_t$ is lower than that of previous case mainly because 
we don't fix $M_H$ at 300 GeV. These values are consistent with recent ones
obtained by the authors of Ref. [3].
The results of this fit are shown in Table 1 as the SM results.

To investigate the effects of the new physics we perform the fit
with the new physics parameters $\kappa_L^b $,
$\kappa_L^c $ and  $\delta \Gamma_s $ fixing $\alpha_s=0.123$. 
This is our new physics fit. We obtain
\begin{eqnarray}
M_H &=& 86^{+ 467}_{-56} ~~{\rm GeV} 
\left[ \log_{10}(M_H)=1.94^{+0.80}_{-0.47} \right], \nonumber \\
M_t &=& 160.9^{+ 28.0}_{- 14.0}, \nonumber \\
\kappa_L^b &=& 0.013 \pm 0.004, \nonumber \\
\kappa_L^c &=& -0.059 \pm 0.026, \nonumber \\
\delta \Gamma_s &=& 18.8 \pm 12.6  ~~ {\rm MeV}. \nonumber 
\end{eqnarray}
As expected $\delta \Gamma_s$ has nearly same value as the deficit of $\Gamma_c$ and is
positive. $\kappa_L^c$ has negative value at 2 $\sigma$ level.
$\kappa_L^b$ has the same central value of our previous work [6] at 3 $\sigma$ level.
$M_t$ is more consistent with the value measured at CDF and D0 than the SM fit is.
The errors of $M_H$ are larger than those of the SM fit
and the center value is higher.
The upper limit at 2 $\sigma$ level is about 2 TeV. This means that  
perturbative calculations are not reliable always. And the
upper limit at 1 $\sigma$ level ($\sim$ 500 GeV) diminishes the hope for
finding the Higgs at the LEP2 or the LHC. In the SM framework, the electroweak
data consistently favor a Higgs mass in a range of ${\cal O}(M_Z)$. 
But, even though it is not significant because of the large error, there is 
a possibility that $M_H$ has a value out of the range of ${\cal O}(M_Z)$.
The results of this fit are shown in Table 1 as the new physics.

To see the effects of future, more precise measurements of $R_b$ and $R_c$ on $M_H$,
we reduce errors of $R_b$ and $R_c$ by half. We do not change the central values 
of $R_b$ and $R_c$. Fixing $\alpha_s=0.123$, we obtain
\begin{eqnarray}
M_H &=& 85 ^{+ 278}_{-59} ~~{\rm GeV} 
\left[ \log_{10}(M_H)=1.93^{+0.63}_{-0.51} \right], \nonumber \\
M_t &=& 160.9^{+ 19.7}_{- 11.8}, \nonumber \\
\kappa_L^b &=& 0.013 \pm 0.002, \nonumber \\
\kappa_L^c &=& -0.063 \pm 0.014, \nonumber \\
\delta \Gamma_s &=& 20.9 \pm 6.7 ~~ {\rm MeV}. \nonumber 
\end{eqnarray}
We observe the errors of $\kappa_L^b$, $\kappa_L^c$, and $\delta \Gamma_s$ decrease.
The errors of $M_H$ decrease slightly and the center value does not change.

To study the effects of fixing $\alpha_s$, we also execute a fit fixing $\alpha_s=0.112$.
We obtain
\begin{eqnarray}
M_H &=& 86 ^{+ 474}_{-55} ~~{\rm GeV} 
\left[ \log_{10}(M_H)=1.94^{+0.81}_{-0.45} \right], \nonumber \\
M_t &=& 160.9^{+ 27.8}_{- 13.9}, \nonumber \\
\kappa_L^b &=& 0.015 \pm 0.004, \nonumber \\
\kappa_L^c &=& -0.057 \pm 0.026, \nonumber \\
\delta \Gamma_s &=& 22.3 \pm 12.6 ~~ {\rm MeV}. \nonumber 
\end{eqnarray}
Comparing with the new physics fit, we can see the effects of fixing $\alpha_s$ 
are negligible.

Because we take a model-independent approach, we do not explicitly describe the parameters
$\kappa_L^b $, $\kappa_L^c $ and  $\delta \Gamma_s $ by specific physical quantities here.
We know, however, that these parameters are related to the new physics scale $\Lambda$.
For example, we consider the t-quark condensation models where the third generation $Q_L$
and $t_R$ states at a minimum participate in a new strong interaction for $\kappa_L^b$ [13]. 
Then the relevant term of the effective Lagrangian is given by
\begin{equation}
{\cal L}_{\rm eff} \sim -\frac{1}{\Lambda^2} \bar{b}\gamma_{\mu}b\bar{t}\gamma^{\mu}
(g_V-g_A\gamma_5)t,
\end{equation}
where $g_V$ and $g_A$ are parameters.
Here one would expect that the $t$-quark loop will generate an effective contribution
to $Z \rightarrow b\bar{b}$ vertex $\kappa_L^b$. Thus we have
\begin{equation}
\kappa_L^b = \frac{g_A}{g_L^b}\frac{N_c}{8 \pi^2}\frac{M_t^2}{\Lambda^2}\ln
\left(\frac{\Lambda^2}{M_t^2}\right),
\end{equation}
where $N_c=3$.
Our fit result $\kappa_L^b = 0.013$ yields $\Lambda \sim 1$ TeV 
with $|g_A| \sim 4\pi(0.11)$ [14].
The results from the analyses of stability [8] and perturbative [9] bounds 
on the SM Higgs boson mass gives
\begin{eqnarray}
\sim 50 ~{\rm GeV}~<~M_H~<~\sim~700~ {\rm GeV}~~~~ {\rm for}~ \Lambda = 1~ {\rm TeV}.
\nonumber
\end{eqnarray}
The perturbative bound 700 GeV gets much corrections from two-loop
$\beta$ functions and one-loop matching condition on the Higgs boson mass. So this
value is considered to be in a range from 500 GeV to 1 TeV. For smaller $\Lambda$
the bounds become weaker. We can see that our new physics fit for $M_H$ is well 
compatible with these bounds.

We implement a fit to LEP observables with new physics explaining 
$R_b$ and $R_c$ discrepancies. 
We obtain $M_H^{\rm New Physics} = 85 ^{+ 467}_{-56} ~~{\rm GeV}$. 
Comparing with the value of the SM fit,
$M_H^{\rm SM} = 38^{+ 96}_{-21} ~~{\rm GeV}$, the errors are larger and
the central value is higher. The new physics may allow $M_H$ to have a
value out of the range of ${\cal O}(M_Z)$.

\section*{Acknowledgments} 
This work was supported in part by Korea Science and Engineering Foundation.  

\newpage


\newpage

\section*{Table Captions}

\begin{description}
\item{Table 1} : Our global fit to LEP observables in the standard model framework
and with nonuniversal interactions explaining $R_b$ and $R_c$ discrepancies.
\end{description}

\newpage


\begin{table}[htb]
\begin{center}
\[
\begin{array}{|c|c|c|c|c|c|}
\hline
&&&&&\\
\mbox{Observables}&\mbox{Experiment}&\mbox{SM results}&~~~\chi^2~~~&\mbox{New Physics }&~~~\chi^2~~~\\
&&&&&\\
\hline
\Gamma_Z (\mbox{GeV})&2.4963 \pm 0.0032&2.4936&0.710&2.4963&0.000 \\
\sigma_{tot} (\mbox{nb})&41.488 \pm 0.078&41.429&0.580&41.441&0.368  \\
R_e&20.797 \pm 0.058&20.799&0.001&20.784&0.052  \\
R_{\mu}&20.796 \pm 0.043&20.799&0.004&20.784&0.079  \\
R_{\tau}&20.813 \pm 0.061&20.846&0.290&20.831&0.087  \\
A_{FB}^0(e)&0.0157 \pm 0.0028&0.0157&0.000&0.0158&0.001  \\
A_{FB}^0(\mu)&0.0163 \pm 0.0016&0.0157&0.134&0.0158&0.103  \\
A_{FB}^0(\tau)&0.0206 \pm 0.0023&0.0157&4.513&0.0158&4.381  \\
A_{\tau}&0.1418 \pm 0.0075&0.1447&0.155&0.1451&0.191  \\
A_{e}&0.139 \pm 0.0089&0.1447&0.417&0.1451&0.466  \\
R_b&~~~0.2219 \pm 0.0017~~~&0.2168&8.868&0.2219&0.000  \\
R_c&0.154 \pm 0.0074&0.1719&5.863&0.1557&0.053  \\
A_{FB}^0(b)&0.0997 \pm 0.0031&0.1016&0.361&0.1020&0.535  \\
A_{FB}^0(c)&0.0729 \pm 0.0058&0.0725&0.006&0.0688&0.494  \\
\sin^2\theta_W^{lep}&0.2320 \pm 0.0016&0.2318&0.014&0.2318&0.021  \\
\hline
\mbox{total}&&&21.9&&6.8  \\
\hline
\end{array}
\]
\end{center}
\caption{
} 
\end{table}

\end{document}